\documentclass[conference]{IEEEtran}

\usepackage[cmex10]{amsmath}
\usepackage{cite}
\usepackage{url}
\usepackage{color}

\usepackage{amssymb}
\usepackage{path}
\usepackage{verbatim}
\usepackage{algorithm}
\usepackage{framed}
\usepackage{algpseudocode}

\usepackage{color}
\usepackage{amsmath}
\usepackage{cases}
\usepackage{theorem}
\usepackage{subfig}
\usepackage{subfloat}

\usepackage{graphicx}

\usepackage{pgfplots}
\usepackage{pgfplotstable}
\pgfplotsset{compat=newest}

{ \theorembodyfont{\normalfont} %\theorembodyfont{\rmfamily}

}

\newcounter{enumctr}

\DeclareFontFamily{U}{mathx}{\hyphenchar\font45}
\DeclareFontShape{U}{mathx}{m}{n}{<-> mathx10}{}
\DeclareSymbolFont{mathx}{U}{mathx}{m}{n}
\DeclareMathAccent{\widebar}{0}{mathx}{"73}

\begin{document}

\title{An Intelligent Multi-Speed Advisory System using Improved Whale Optimisation Algorithm}

\author{\IEEEauthorblockN{Beiran Chen \IEEEauthorrefmark{1},
		Mingming Liu \IEEEauthorrefmark{2}\IEEEauthorrefmark{4}, Yi Zhang \IEEEauthorrefmark{3}, Zhengyong Chen\IEEEauthorrefmark{2}, Yingqi Gu \IEEEauthorrefmark{2}  and Noel E. O'Connor \IEEEauthorrefmark{2}}
	    \IEEEauthorblockA{\IEEEauthorrefmark{1}  CONNECT centre, Trinity College Dublin, Ireland \\
		\IEEEauthorrefmark{2} School of Electronic Engineering, Insight Centre for Data Analytics, Dublin City University, Ireland \\
		\IEEEauthorrefmark{3} Huawei Technologies Ireland. \\
		\textit{\IEEEauthorrefmark{4} Joint first author \& corresponding author, Email}: mingming.liu@dcu.ie}}

% make the title area
\maketitle

%\textcolor[rgb]{1,0,0}{

\begin{abstract}

An intelligent speed advisory system can be used to recommend speed for vehicles travelling in a given road network in cities. In this paper, we extend our previous work where a distributed speed advisory system has been devised to recommend an optimal consensus speed for a fleet of Internal Combustion Engine Vehicles (ICEVs) in a highway scenario. In particular, we propose a novel optimisation framework where the exact format of each vehicle's cost function can be implicit, and our algorithm can be used to recommend multiple consensus speeds for vehicles travelling on different lanes in an urban highway scenario. Our studies show that the proposed scheme based on an improved whale optimisation algorithm can effectively reduce $CO_2$ emission generated from ICEVs while providing different recommended speed options for groups of vehicles.

\end{abstract}

% Note that keywords are not normally used for peerreview papers.

\IEEEpeerreviewmaketitle

\section{Introduction}

An Intelligent Speed Advisory (ISA) system, as part of the Advanced Driver Assistant System (ADAS), has been widely used to provide drivers with useful speed information in travel \cite{paul2016advanced, wan2016optimal, martinez2017driving, yue2018assessment}. The provided information to drivers can be effectively used to improve drivers' safety and experience, reduce traffic time and congestions, as well as to improve vehicles' dynamic performance and efficiency, among others \cite{lu2005technical, gruyer2011distributed, liu2017fine, wang2020digital, gamez2017dynamic, xiang2015closed}. In our previous work, we have explored the design of a Distributed Speed Advisory System (DSAS) where an optimal consensus recommended speed can be sought so that some specific objectives can be achieved for smart transportation applications. This includes minimisation of the overall $CO_2$ emissions for a fleet of Internal Combustion Engine Vehicles (ICEVs) in a highway scenario \cite{gu2014optimised, liu2015distributed}, minimisation of the overall energy consumption for a group of electric vehicles \cite{liu2015intelligent, liu2015distributed}, and maximisation of the overall health benefits for a cycling group involving e-bikes (electric bikes) \cite{gu2018design}.

In this follow-on work, our main objective is still to design a DSAS in order to reduce the overall emission levels for a group of ICEVs. However, instead of having the prior knowledge of the exact mathematical expression of each vehicle's cost function, which we assumed in \cite{liu2015distributed}, here we explicitly assume that this information can be implicit in the sense that each local vehicle may not easily get access to such information in a practical scenario. Our starting point is the observation that an average speed model used in \cite{liu2015distributed} may not perfectly represent the emission generation of a vehicle in a practical scenario. For instance, the averaged emission generated by a vehicle travelling at $70$km/h on a flat road may not be the same compared to the vehicle driving on a sloped road at the same speed. In fact, real emission generation of a given vehicle can be measured by sensors and this raw data can be collected, transmitted to the cloud for further data processing and analysis by Artificial Intelligence (AI) and machine learning related techniques. In this context, an AI-based prediction model is preferable for providing an accurate evaluation compared to conventional mathematical models. We note that the prediction model can be trained either on the cloud or on the local side, but in either case the prediction model is usually seen as a ``black-box'' to users.

Our second key observation is that a single recommended consensus speed may not be sufficient for a practical highway scenario involving multi-lane roads. In such a case, the recommended speed may be even lower than the expected speed of a driver as the optimal speed depends on the composition of the fleet. On the other side, a single recommended speed may not always be desirable from the perspective of a highway operator as insufficient usage of road capacity can significantly reduce traffic flow of road networks. Therefore, it is important to allow a driver to opt-in to a different ``optimal speed'' when such a request can create a positive impact on both drivers and the road networks. 

Along this line, our key objective in this paper is to devise a DSAS such that different optimal speeds can be recommended to different lanes on a highway, and in such a way that when different groups of vehicles following their respective optimal speed, the overall $CO_2$ emission of vehicles can be minimised for the road networks. In particular, we shall assume that the emission generation of a given ICEV can only be evaluated using a ``black-box'' model, which implies that a given vehicle can only get access to very limited information of its emission function, e.g. using REST API calls to the cloud associated with a cost, but with no access to derivative-related information.

We note that in this presented problem of interest, traditional derivative-based optimisation methods, e.g. gradient descent, are no longer suitable as derivative-related information cannot be retrieved for algorithm implementations.  In this regard, a centralized-based heuristic algorithm may be plausible for an optimal solution, but such an algorithm usually needs to be deployed on a central server. This typically means a base station which requires to collect full information, i.e. emission functions, from vehicles for the calculation, which may not be practical in scenarios where these function can only be evaluated locally or through an encrypted private channel to the cloud. Thus, a distributed design of SAS, i.e. DSAS, is more preferable. 

Given these considerations, the contribution of this paper is to propose a new design solution for the DSAS by including the following features:

\begin{itemize}
	\item  An intelligent distributed optimisation algorithm that can be implemented for the DSAS with limited access to the emission functions, i.e. derivative-free.
	\item  An improved heuristic optimisation algorithm implementation that can easily deal with proportional-based consensus constraints and can converge rapidly compared to other centralized-based heuristic algorithms. 
	\item  A DSAS which can recommend different optimal speeds to multi-lanes of vehicles with users'  preference included.
	\item  A comprehensive evaluation of experiments showing how users' preference can affect the overall optimal emission target in various highway scenarios. 
\end{itemize}

We believe that the benefits of our proposed solutions are particularly appealing for traffic scenarios where:

\begin{itemize}
	\item  a highway operator who wishes to provide a speed advisory service for vehicles on its road networks where both $CO_2$ emission target and the drivers' preference for speeds are equivalently important.
	\item  users' privacy is important in the algorithm implementation for optimality. 
	\item  existing traffic infrastructures are sufficient to satisfy the requirements of the system/application implementation. 
\end{itemize}

The paper is organised as follows. In Section \ref{sec:System_model}, we introduce the system model for the problem of interest as a utility optimisation problem with proportional consensus constraints. In Section \ref{sec:optalg}, we give a quick recap of the Whale Optimisation Algorithm and then we show that how the algorithm can be further adapted and improved for solving our problem.  In Section \ref{sec:Experimental_Results}, we demonstrate the simulation setup and illustrate some experimental results.  Finally, we conclude the paper in Section \ref{sec:Conclusion} and outline some future scope of the paper.

\section{System Model}\label{sec:System_model}

%In this section, we present our system model for the optimisation problem. 

Let us consider a scenario in which a number of ICEVs are driving along a stretch of highway in different lanes. We assume that there are $N$ drivers of ICEVs that are willing to use the speed advisory system during their driving period on the highway. Let $\underbar{N}:=\left\{1,2,...,N\right\}$ be the set for indexing the ICEVs, and let $s_i(k)$ denote the recommended speed of the $i^{th}$ vehicle at a time slot $k$. In addition, let $\textbf{s}(k)^{T}:= \left[s_1(k), s_2(k), \ldots, s_N(k)\right]$ be the vector of the recommended speeds of all ICEVs. Let $\alpha_i$ be the recommended speed parameter which indicates the level of speed that a driver expects to achieve compared to other vehicles. For instance, one vehicle can set the parameter to 0.5 to indicate its expected speed is twice to another vehicle which sets this parameter value to 1. In addition, let $\underbar{s}_i$ and $\overline{s}_i$ denote the lower and upper bound of the recommended speed that the $i^{th}$ vehicle can follow respectively. Furthermore, each ICEV $i$ is associated with a cost function $f_i(s_i(k))$, a black-box function, which returns the amount of $CO_2$ emission generated by the $i^{th}$ vehicle at its current state with respect to the current recommended speed $s_i(k)$. In comparison to our previous work \cite{liu2015distributed}, this function as used in this paper does not need to be strict convex and differentiable. We shall only require that the function can be evaluated for a given range of speed. With this in place, the specific problem to be solved can be formulated as follows:

%For this purpose, we assume that each vehicle is able to receive the recommended speed from a road infrastructure, transmit limited information back to the infrastructure, and the vehicle is also able to send a broadcast message to its nearby vehicles. 

\begin{equation} \label{eq:opt}
\begin{gathered}
\underset{s_1, s_2, \ldots, s_{N}}{\min} \quad
\sum\limits_{i\in\underbar{N}} f_{i}\left(s_i \right),\\
{\text{s.t.}} ~
s_i \in [\underbar{s}_i, \overline{s}_i],  ~\alpha_i s_i = \alpha_j s_j, ~ \forall i, j \in \underbar{N}.
\end{gathered}
\end{equation}

\begin{figure*}[ht]
	\vspace{-0.1in}
	\centering
	\includegraphics[width=0.63\textwidth, height=3.5in]{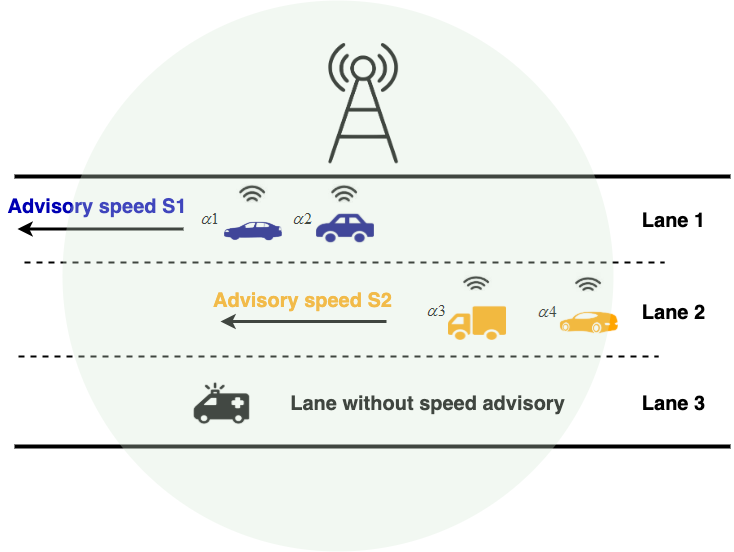}
	\caption{Schematic diagram of the proposed system architecture.}
	\label{fig:schematic}
	\vspace{-0.1in}
\end{figure*}

\noindent \textbf{Comment:} The options for $a_i$ should be limited in a practical scenario. For instance, in a 3 lane highway where 2 lanes are running the speed advisory system, $a_i$ can have two options: one large value indicates an expected ``slow'' recommended speed and one small value indicates an expected  ``fast'' recommended speed. The hope is that by using an iterative algorithm, which we shall introduce in the following section, $s_i(k)$ can converge to the optimal solution which addresses \eqref{eq:opt}. To illustrate this fundamental concept, a schematic diagram of the system model is shown in Fig. \ref{fig:schematic}, where the two blue vehicles are driving under the same recommended speed, the two yellow cars are driving under another common recommended speed, and the vehicle on the Lane 3 can be running without any speed advisory. In a real world scenario, the Lane 3 can be used for any emergency vehicles, e.g. police vehicles or ambulances, where a recommended speed is not primarily important.

\section{Optimisation Algorithm} \label{sec:optalg}

In this section, we briefly review the Whale Optimisation Algorithm (WOA) and then we show that how WOA can be adapted to solve the optimisation problem presented in \eqref{eq:opt}.

\subsection{Whale Optimisation Algorithm} \label{WOA}

The Whale Optimisation Algorithm was initially proposed in \cite{mirjalili2016whale} to solve complex ``black-box'' optimisation problems using meta-heuristic. The algorithm is inspired by the hunting behaviour of humpback whales. The WOA implements three steps, namely search for prey, encircling prey, and bubble-net attacking. At a given time slot, a whale can either encircling a prey or attack the prey using bubble-net, which is done in a random manner. The key idea behind is to gradually reduce the search space in a way that the prey, which is the best candidate solution of the optimisation problem, can be updated along the movement of the whales, i.e. search agents. Now we present a brief overview of the original WOA \cite{mirjalili2016whale} which includes the three key steps mentioned. \\

\subsubsection{Encircling prey}

In this step, whales can recognize the location of the pray and encircle them by updating the position vector $\textbf{X}(t)$ as follows:

\begin{equation} \label{ep1}
\textbf{D} = | \textbf{C} \cdot \textbf{X}^{*}(t) - \textbf{X}(t)| 
\end{equation}
\begin{equation} \label{ep2}
\textbf{X}(t+1) =   \textbf{X}^{*}(t) - \textbf{A} \cdot \textbf{D} 
\end{equation}

\noindent where $\textbf{D}, \textbf{C}, \textbf{A}, \textbf{X}^{*}(t)$ are also vectors. The dimension of each vector is same and is determined by the number of variables to be solved. ``.'' denotes element-by-element multiplication. $\textbf{X}^{*}(t)$ denotes the position of the prey and thus the position of the optimal solution. In particular, $\textbf{A}$ and $\textbf{C}$ are updated as follows:

\begin{equation} \label{defa}
\textbf{A} = 2\alpha \cdot \textbf{r} - \alpha
\end{equation}
\begin{equation} \label{upc}
\textbf{C} = 2 \cdot \textbf{r}
\end{equation}

\noindent where $\alpha$ is a scalar linearly decreased from 2 to 0 during algorithm iterations, and $\textbf{r}$ is a random vector in $[0, 1]$.

\subsubsection{Bubble-net Attacking}

This step models the behaviour of a whale when attacking a prey. The whale will swim along a spiral path to create distinctive bubbles. The position of the whale is updated using the following equation. 

\begin{equation} \label{bna}
\textbf{X}(t+1) = | \textbf{X}^{*}(t) - \textbf{X}(t) | \cdot e^{bl} \cdot cos(2 \pi l) + \textbf{X}^{*}(t)
\end{equation}

\noindent where $b$ is a constant (chosen 1 by default), and $l$ is a random number between $[-1, 1]$.  \\

\subsubsection{Searching for Prey}

As we mentioned earlier, at every given time instance $t$, a whale is either attacking a prey using bubble-net or encircling prey by updating its position. This decision is made in a fully random manner, i.e. $50\%$ probability for each behaviour. If a whale's decision is to encircle the prey, a new position will be updated according to the value of $|\textbf{A}|$. We note that according to the definition of $\textbf{A}$ in \eqref{defa}, each element of $\textbf{A}$ is a random number between $[-\alpha, \alpha]$, where $\alpha$ also linearly decreases from 2 to 0 during the algorithm iterations. Finally, the position of a whale is updated using \eqref{ep1} and \eqref{ep2} if $|\textbf{A}| < 1$, and using \eqref{sfp1} and \eqref{sfp2} if $|\textbf{A}| \geq 1$.

\begin{equation} \label{sfp1}
\textbf{D} = | \textbf{C} \cdot \textbf{X}_{\textrm{rand}}(t) - \textbf{X}(t)| 
\end{equation}
\begin{equation} \label{sfp2}
\textbf{X}(t+1) =   \textbf{X}_{\textrm{rand}}(t) - \textbf{A} \cdot \textbf{D}. 
\end{equation}

\noindent  In the above formulas,  $\textbf{X}_{\textrm{rand}}(t)$ denotes a random position of a whale in the population. The pseudocode of original WOA \cite{mirjalili2016whale} is presented in Algorithm \ref{woa_sim}. \\

% \cite{mirjalili2016whale}.\MMcom{Maybe good to put the equations from the original paper? Also, it may be good to explain a bit about how the WOA algorithm can be used to solve the constrained optimization problem, using different penalty functions, in a centralized framework. } \\

\begin{algorithm}[htbp]
	\caption{WOA \cite{mirjalili2016whale}}
	\begin{algorithmic}[1]
		\State Initialise the whales population $\textbf{X}_i, ~ (i = 1,2, \dots, n)$. 
		\State Calculate the fitness of each search agent and identify the best search agent $\textbf{X}^{*}$.
		\While{$k <$ maximum number of iterations}
		\For{each search agent}
		\State Update $\textbf{A}$, $\textbf{C}$ using \eqref{defa} and \eqref{upc}.
		\State Update random numbers $l \in [-1, 1]$ and $p \in [0,1]$. 
		\If{$p < 0.5$}
		\If{$|\textbf{A}| < 1$}
		\State Update the position of current search agent using \eqref{ep1} and \eqref{ep2}.
		\ElsIf{$|\textbf{A}| \geq 1$} 
		\State  Select a random search agent ($\textbf{X}_{\textrm{rand}}$).
		\State  Update the position of current search agent using \eqref{sfp1} and \eqref{sfp2}. 
		\EndIf           
		\ElsIf{$p \geq 0.5$}
		\State Update the position of current search agent using \eqref{bna}.
		\EndIf
		\EndFor
		\State  Check if any search agent goes beyond the search space and amend it.
		\State  Calculate the fitness of each search agent. 
		\State  Update $\textbf{X}^{*}$ if there is a better solution. 
		\State  $k = k + 1$
		\EndWhile
		\State   Return $\textbf{X}^{*}$. 
	\end{algorithmic}
	\label{woa_sim}
\end{algorithm}

\noindent \textbf{Comment:} 

\begin{itemize}	
	\item The original WOA  \cite{mirjalili2016whale} can be easily implemented using a centralized based architecture, where each search agent can be seen as an independent process of a central computing node. In our problem, each search agent can represent a vector of recommended speed $[s_1, s_2, \ldots, s_N]$, and a central computing node needs to initialise a population of search agents to find the optimal $\textbf{s}(k)^{T}$ from a group of search agents at the end of the algorithm iterations. 	
	\item The original WOA \cite{mirjalili2016whale} intends to find out the best search agent based on a fitness function which depends on the states of all search agents. In this context, the fitness function requires information of all cost functions of all ICEVs, and this implies that each ICEV user needs to reveal its cost function $f_i$ to a central computing node, e.g. a roadside base station, which on the one side may not be possible (black-box models) and on the other side it can also be of privacy concerns to users.
\end{itemize}

\subsection{The Improved WOA for Multi-speed Advisory System}

In the literature, WOA has shown to be an effective algorithm for solving challenging black-box optimisation problems \cite{mirjalili2016whale}. However, the original WOA in \cite{mirjalili2016whale} needs to be implemented in a centralised manner, where in our context all information from users needs to be collected, including the recommended speed parameters and the cost functions, for decision making which may not be desirable to users. Instead, we borrow the fundamental ideas of WOA and improve it so that the adapted WOA can be implemented in a distributed framework. The key benefit of this approach is that drivers can deploy trained models to evaluate emission without revealing such information to a central computing node, e.g. road infrastructure. Our improved WOA is presented in Algorithm \ref{woa_new}.

%This is well aligned with the privacy-preserving feature proposed in \cite{liu2015distributed}. Our improved WOA is presented in Algorithm \ref{woa_new}.

\begin{algorithm}[htbp]
	\caption{Improved WOA for the proposed system}
	\begin{algorithmic}[1]
		
		\State Each ICEV $i \in \underline{\textrm{N}}$ initialises $M$ whales in a sequence, i.e. $s_i^{1}(k), s_i^{2}(k), \dots, s_i^{h}(k), \dots, s_i^{M}(k)$. 
		
		\State Initialisation and synchronisation of the parameters $a, b > 0 $ in the vehicular network. 
		
		\State Each ICEV $i$ evaluates $af_i(s_i^{h}(k))+b$ for $h=1,2,\dots, M$, and sends these $M$ values to a central node in a sequence. 
		
		\State The central node aggregates $\sum_{i \in \underline{\textrm{N}}} af_i(s_i^{h}(k))+b, \forall h$ in a sequence, and finds the index $h^{*}$ which gives rise to $h^{*} = \underset{h}{\mathrm{argmin}} \sum_{i \in \underline{\textrm{N}}} af_i(s_i^{h}(k))+b, ~\forall h \in \{1,2,...,M\}$.
		
		\State The central node broadcasts $h^{*}$ to all  ICEVs. 
		
		\While{$k < k_{\textrm{max}}$}
		
		\State Use WOA \cite{mirjalili2016whale} to update $s_j^{h}(k+1) \leftarrow s_j^{h}(k)$, $\forall h$ for a random ICEV $j \in \underline{\textrm{N}}$.
		
		\State  Check if $s_j^{h}(k+1) \in [\underbar{s}_i, \overline{s}_i], \forall h$, if not then bound it.
		\State  ICEV $j$ broadcasts $\alpha_j s_j^{h}(k+1), \forall h$ to ICEV $i \in \underline{\textrm{N}}$. 
		\State  Update $s_i^{h}(k+1)$ to  $\alpha_j s_j^{h}(k+1)/\alpha_i$, $\forall i, ~\forall h$.
		\State  ICEV $i$ evaluates $af_i(s_i^{h}(k+1))+b, \forall i, h$ and sends them to the central node.
		\State  Central node aggregates $\sum_{i=1}^{N} af_i(c_i^{h}(k+1))+b, \forall h$, and updates $h^{*}$ if it results in a smaller value. 
		\State  The central node broadcasts $h^{*}$ to all ICEVs. 
		\State  $k = k + 1$
		\EndWhile 
		\State   Return $s_{i}^{h^{*}}(k_{\textrm{max}}), \forall i \in \underline{\textrm{N}}$. 
	\end{algorithmic}
	\label{woa_new}
\end{algorithm}

\begin{figure}[ht]
	\vspace{-0.1in}
	\centering
	\includegraphics[width=0.47\textwidth, height=3.2in]{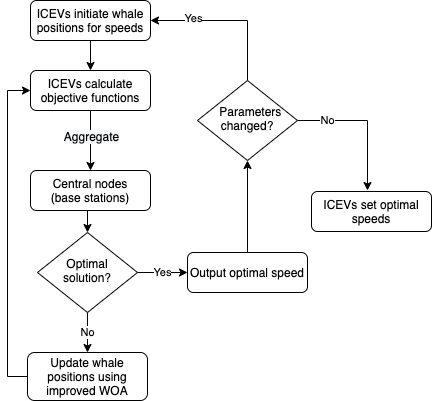}
	\caption{Flow chart of the proposed system implementation.}
	\label{fig:flow chart}
	\vspace{-0.1in}
\end{figure}

With the improved WOA in place, our proposed system is implemented in the following 6 key steps, which are illustrated in the rectangle boxes in the flow chart in Fig. \ref{fig:flow chart}. It is worth noting that after an optimal speed has been found by the improved WOA, the system needs to regularly check if any system parameters have been changed, e.g. $\alpha_i, \alpha_j$, and if so the system needs to reactivate the WOA by finding the new optimal speeds for the group of ICEVs. This feature makes the system adaptive as it can find the optimal solution in a timely, online and dynamic manner. 

\section{Experimental Results}\label{sec:Experimental_Results}

\begin{figure}[ht]
	\vspace{-0.1in}
	\centering
	\includegraphics[width=0.5\textwidth, height=2.3in]{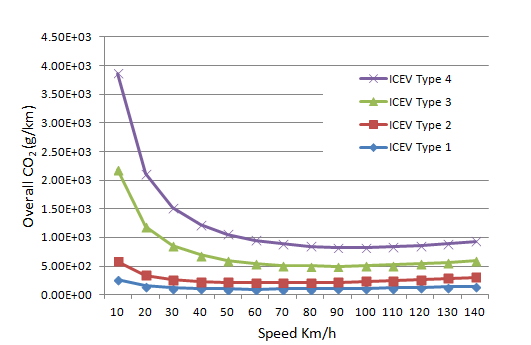}
	\caption{$CO_2$ emission models for different types of ICEVs.}
	\label{fig:emission}
	\vspace{-0.1in}
\end{figure}

We use Matlab and the open-source whale optimisation software in \cite{mirjalili2016whale} to evaluate the performance of the improved WOA. To begin with, we consider a typical highway scenario with two lanes running the proposed WOA, corresponding to the fast and slow lanes of a highway, and we assume that the speed range of the highway is [60, 120] km/h. For each speed advisory there are three whales initialised for the WOA, and the objective is to minimise the total $CO_{2}$ emission of vehicles on the highway where 4 types of ICEVs are presented in Fig. \ref{fig:emission} with 10 vehicles each for Type 1 and 4, and 20 vehicles each for Type 2 and 3.

\begin{figure}[ht]
	\vspace{-0.1in}
	\centering
	\includegraphics[width=0.6\textwidth, height=2.5in]{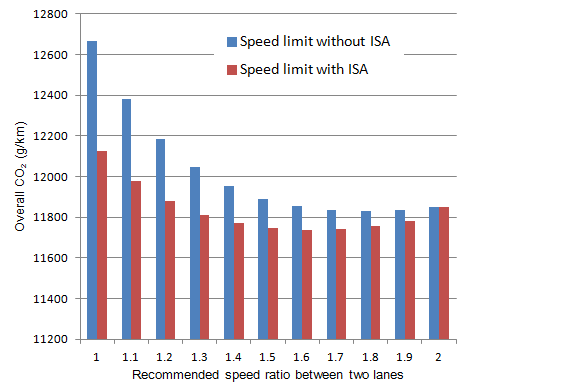}
	\caption{Evaluation of the overall $CO_2$ emissions generated by vehicles on two lanes with and without the ISA implemented. }
	\label{fig:iteration_steps}
	\vspace{-0.1in}
\end{figure}

\begin{figure}[ht]
    \vspace{-0.1in}
	\centering
\includegraphics[width=0.47\textwidth, height=2.5in]{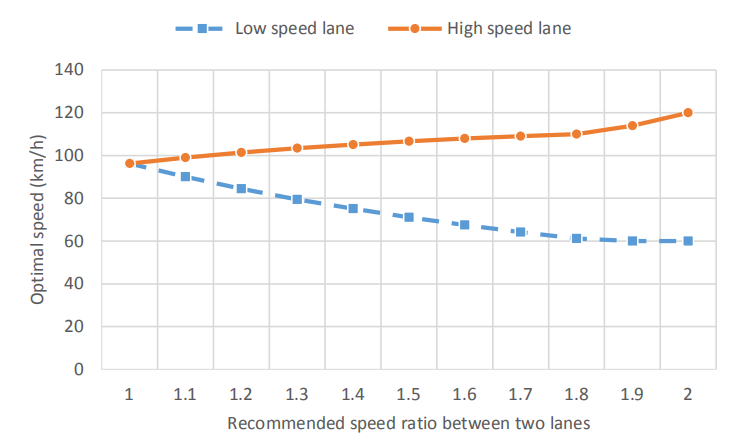}
	\caption{Optimal recommended speeds of vehicles on two lanes with respect to the varying desired recommended speed ratio. }
	\label{fig:optimal_speed_ratio}
    \vspace{-0.1in}
\end{figure}

To illustrate the benefits of the proposed system, we evaluated the overall $CO_2$ emissions of vehicles on both lanes with respect to the changing value of the desired recommended speed ratio with and without the proposed ISA. For the setup, we assumed that there are 10 Type-1 vehicles and 20 Type-2 vehicles on the slow lane, and 20 Type-3 and 10 Type-4 vehicles driving on the fast lane. We note that since the speed range of the highway is [60, 120] km/h, the maximum recommended speed ratio can only be 2 in our simulated scenarios. Here, the baseline setup (i.e. without the proposed ISA) assumes a greedy speed requirement, where the vehicles on the fast lane will always drive at the speed limit, i.e. 120 km/h, and the speed of vehicles on the other lane will be subject to the ratio of recommended speed parameters. Our simulation results are illustrated in Fig. \ref{fig:iteration_steps} which shows that with the increasing ratio of the recommended speed parameter between two lanes, a decreasing $CO_2$ saving has been found when compared to the baseline setup, with the highest saving being 543 g/km when all lanes achieve a common recommended speed.  Interestingly, we also find that the $CO_2$ saving vanishes when the ratio becomes 2, corresponding to the only feasible solution 60 km/h (slow lane) and 120 km/h (fast lane) with/without the ISA applied. In addition, the optimal recommended speeds of vehicles on both lanes are shown in Fig. \ref{fig:optimal_speed_ratio}. The horizontal axis shows the ratio increasing from 1 to 2 at which point the speed limit 120 km/h is reached. Clearly, results in both figures together reveal the trade-off between users' flexibility in speed and the $CO_2$ saving at an aggregated level.

\begin{figure}[ht]
	\vspace{-0.1in}
	\centering
	\includegraphics[width=0.5\textwidth, height=2.5in]{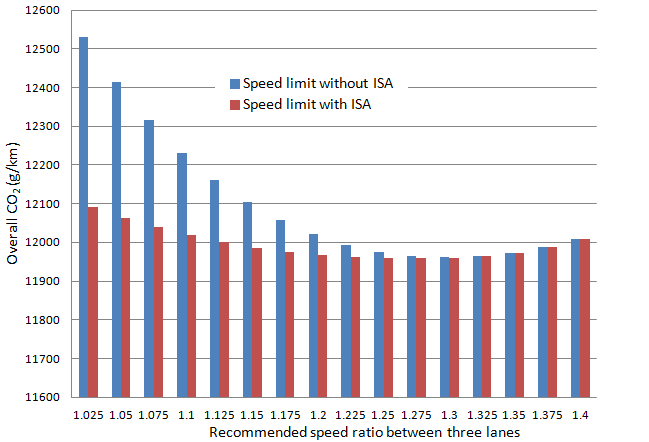}
	\caption{Overall $CO_2$ saving with respect to recommended speed ratio between three lanes.}
	\label{fig:co2_3lanes}
	\vspace{-0.1in}
\end{figure}

To further illustrate the insight of the system, we evaluated the overall $CO_2$ emissions of the same group of vehicles on a three-lane-highway. For simplicity, we assumed that the recommended speed ratio between the first and second lane is the same as the ratio between second and the third lane. This leads to a three-lane-highway scenario, representing slow/medium/fast-speed lanes. As a result, the recommended speed ratio is only available in the range [1, $\sqrt{2}$]. We also assumed that there are 10 Type-1 vehicles on the slow lane, 20 Type-2 and 20 Type-3 vehicles on the medium lane, as well as 10 Type-4 vehicles on the fast lane. The simulation results are illustrated in Fig. \ref{fig:co2_3lanes}. Similarly to the two-lane-highway scenario, the results show that with the increasing ratio of the recommended speed parameter between  three  lanes,  a  decreasing $CO_2$ has been  found with  the highest saving occurs at 440 g/km when all  lanes achieve a close recommended speed. It has been found that with less flexibility of the recommended speed ratio, the benefits of the $CO_2$ saving vanishes rapidly in this three-lane case, and when the ratio is greater than 1.3 the $CO_2$ saving is almost negligible.

\begin{figure}[ht]
	\vspace{-0.1in}
	\centering
	\includegraphics[width=0.5\textwidth, height=2.5in]{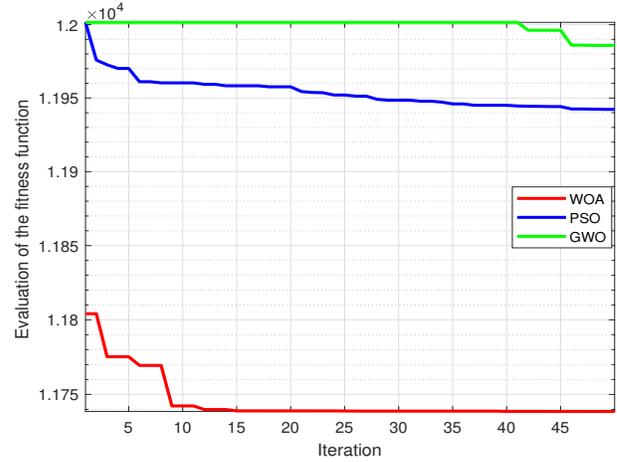}
	\caption{Evolution of the fitness function value of the improved WOA compared with two baseline algorithms.}
	\label{fig:alg_compare}
	\vspace{-0.1in}
\end{figure}

Finally, we conducted a comparative simulation for the improved WOA by comparing its performance with other baseline heuristic methods including Particle Swarm Optimisation (PSO) \cite{poli2007particle} and Grey Wolf Optimisation (GWO) \cite{mirjalili2014grey}. In order to deal with the constraints for the optimizers, a death penalty has been applied to both PSO and GWO methods where the penalty value has been set to 100 indicating a penalty will be added if the searching agents cannot achieve the required consensus condition. Our results are illustrated in Fig. \ref{fig:alg_compare}. The results illustrate that GWO has the slowest convergence rate compared to both PSO and the improved WOA, however both PSO and GWO have not been able to converge to the desired optimal solution within the 50 iterations. In contrast, the proposed WOA is able to converge rapidly within 15 steps which clearly demonstrates its merit for the system efficiency.

%\begin{figure}[ht]
%	\vspace{-0.1in}
%	\centering
%	\includegraphics[width=0.45\textwidth, height=2.6in]{co2_saving}
%	\caption{Overall $CO_2$ saving with respect to recommended speed ratio between two lanes}
%	\label{fig:co2_saving}
%	\vspace{-0.1in}
%\end{figure}

\section{Conclusion}\label{sec:Conclusion}

In this paper, we propose an improved WOA algorithm for the new design of an intelligent ISA which can be used effectively to reduce $CO_2$ emissions on multi-lane highways. In our future work, we will investigate how different drivers' behaviour models affect system performance in a real-time testing environment. 

\section*{Acknowledgement}

This material is based on works supported by Science Foundation Ireland under Grant SFI/12/RC/2289\_P2.

\bibliographystyle{ieeetran}
\bibliography{References}

% Generated by IEEEtran.bst, version: 1.14 (2015/08/26)
\begin{thebibliography}{10}
\providecommand{\url}[1]{#1}
\csname url@samestyle\endcsname
\providecommand{\newblock}{\relax}
\providecommand{\bibinfo}[2]{#2}
\providecommand{\BIBentrySTDinterwordspacing}{\spaceskip=0pt\relax}
\providecommand{\BIBentryALTinterwordstretchfactor}{4}
\providecommand{\BIBentryALTinterwordspacing}{\spaceskip=\fontdimen2\font plus
\BIBentryALTinterwordstretchfactor\fontdimen3\font minus
  \fontdimen4\font\relax}
\providecommand{\BIBforeignlanguage}[2]{{%
\expandafter\ifx\csname l@#1\endcsname\relax
\typeout{** WARNING: IEEEtran.bst: No hyphenation pattern has been}%
\typeout{** loaded for the language `#1'. Using the pattern for}%
\typeout{** the default language instead.}%
\else
\language=\csname l@#1\endcsname
\fi
#2}}
\providecommand{\BIBdecl}{\relax}
\BIBdecl

\bibitem{paul2016advanced}
A.~Paul, R.~Chauhan, R.~Srivastava, and M.~Baruah, ``Advanced driver assistance
  systems,'' SAE Technical Paper, Tech. Rep., 2016.

\bibitem{wan2016optimal}
N.~Wan, A.~Vahidi, and A.~Luckow, ``Optimal speed advisory for connected
  vehicles in arterial roads and the impact on mixed traffic,''
  \emph{Transportation Research Part C: Emerging Technologies}, vol.~69, pp.
  548--563, 2016.

\bibitem{martinez2017driving}
C.~M. Martinez, M.~Heucke, F.-Y. Wang, B.~Gao, and D.~Cao, ``Driving style
  recognition for intelligent vehicle control and advanced driver assistance: A
  survey,'' \emph{IEEE Transactions on Intelligent Transportation Systems},
  vol.~19, no.~3, pp. 666--676, 2017.

\bibitem{yue2018assessment}
L.~Yue, M.~Abdel-Aty, Y.~Wu, and L.~Wang, ``Assessment of the safety benefits
  of vehicles’ advanced driver assistance, connectivity and low level
  automation systems,'' \emph{Accident Analysis \& Prevention}, vol. 117, pp.
  55--64, 2018.

\bibitem{lu2005technical}
M.~Lu, K.~Wevers, and R.~Van Der~Heijden, ``Technical feasibility of advanced
  driver assistance systems (adas) for road traffic safety,''
  \emph{Transportation Planning and Technology}, vol.~28, no.~3, pp. 167--187,
  2005.

\bibitem{gruyer2011distributed}
D.~Gruyer, S.~Glaser, S.~Pechberti, R.~Gallen, and N.~Hautiere, ``Distributed
  simulation architecture for the design of cooperative adas,'' in \emph{First
  International Symposium on Future Active Safety Technology toward
  zero-traffic-accident}, 2011.

\bibitem{liu2017fine}
H.~Liu, H.~Wei, T.~Zuo, Z.~Li, and Y.~J. Yang, ``Fine-tuning adas algorithm
  parameters for optimizing traffic safety and mobility in connected vehicle
  environment,'' \emph{Transportation research part C: emerging technologies},
  vol.~76, pp. 132--149, 2017.

\bibitem{wang2020digital}
Z.~Wang, X.~Liao, X.~Zhao, K.~Han, P.~Tiwari, M.~J. Barth, and G.~Wu, ``A
  digital twin paradigm: Vehicle-to-cloud based advanced driver assistance
  systems,'' in \emph{2020 IEEE 91st Vehicular Technology Conference
  (VTC2020-Spring)}.\hskip 1em plus 0.5em minus 0.4em\relax IEEE, 2020, pp.
  1--6.

\bibitem{gamez2017dynamic}
C.~G{\'a}mez~Serna and Y.~Ruichek, ``Dynamic speed adaptation for path tracking
  based on curvature information and speed limits,'' \emph{Sensors}, vol.~17,
  no.~6, p. 1383, 2017.

\bibitem{xiang2015closed}
X.~Xiang, K.~Zhou, W.-B. Zhang, W.~Qin, and Q.~Mao, ``A closed-loop speed
  advisory model with driver's behavior adaptability for eco-driving,''
  \emph{IEEE Transactions on Intelligent Transportation Systems}, vol.~16,
  no.~6, pp. 3313--3324, 2015.

\bibitem{gu2014optimised}
Y.~Gu, M.~Liu, E.~Crisostomi, and R.~Shorten, ``Optimised consensus for highway
  speed limits via intelligent speed advisory systems,'' in \emph{2014
  International Conference on Connected Vehicles and Expo (ICCVE)}.\hskip 1em
  plus 0.5em minus 0.4em\relax IEEE, 2014, pp. 1052--1053.

\bibitem{liu2015distributed}
M.~Liu, R.~H. Ord{\'o}{\~n}ez-Hurtado, F.~Wirth, Y.~Gu, E.~Crisostomi, and
  R.~Shorten, ``A distributed and privacy-aware speed advisory system for
  optimizing conventional and electric vehicle networks,'' \emph{IEEE
  Transactions on Intelligent Transportation Systems}, vol.~17, no.~5, pp.
  1308--1318, 2015.

\bibitem{liu2015intelligent}
M.~Liu, R.~H. Ord{\'o}{\~n}ez-Hurtado, F.~R. Wirth, Y.~Gu, E.~Crisostomi, and
  R.~Shorten, ``An intelligent speed advisory system for electric vehicles,''
  in \emph{2015 International Conference on Connected Vehicles and Expo
  (ICCVE)}.\hskip 1em plus 0.5em minus 0.4em\relax IEEE, 2015, pp. 84--88.

\bibitem{gu2018design}
Y.~Gu, M.~Liu, M.~Souza, and R.~N. Shorten, ``On the design of an intelligent
  speed advisory system for cyclists,'' in \emph{2018 21st International
  Conference on Intelligent Transportation Systems (ITSC)}.\hskip 1em plus
  0.5em minus 0.4em\relax IEEE, 2018, pp. 3892--3897.

\bibitem{mirjalili2016whale}
S.~Mirjalili and A.~Lewis, ``The whale optimization algorithm,'' \emph{Advances
  in engineering software}, vol.~95, pp. 51--67, 2016.

\bibitem{poli2007particle}
R.~Poli, J.~Kennedy, and T.~Blackwell, ``Particle swarm optimization,''
  \emph{Swarm intelligence}, vol.~1, no.~1, pp. 33--57, 2007.

\bibitem{mirjalili2014grey}
S.~Mirjalili, S.~M. Mirjalili, and A.~Lewis, ``Grey wolf optimizer,''
  \emph{Advances in engineering software}, vol.~69, pp. 46--61, 2014.

\end{thebibliography}

\end{document}